\documentclass[aps,prl,twocolumn,showpacs]{revtex4}
\usepackage{graphicx}
\usepackage{dcolumn}
\usepackage{amsmath}
\usepackage[latin1]{inputenc}

\begin{document}

\title[Aging  and SOC in driven dissipative systems]{Aging  and SOC in driven dissipative systems}
\author{Paolo Sibani}\email[Corresponding author ]{paolo@planck.fys.ou.dk}
\author{Christian Maar Andersen}
\affiliation{Fysisk Institut,  
SDU-Odense Universitet, Denmark}

\date{\today}

\begin{abstract}
We study the noisy dynamics 
of a close relative 
to the sand pile model.
Depending on the type of noise  and the time scale of 
observation, we   
find   stationary fluctuations (similar to SOC)
or an aging dynamics with
punctuated equilibria, a decreasing rate of 
events and reset properties  qualitatively similar
to those of  glassy systems, evolution models and 
vibrated granular media. 
\end{abstract}



\maketitle

\section{Introduction}

The  `pulse-duration memory effect' 
observed~\cite{Fleming86}
in sliding Charge Density Wave systems 
was explained by Coppersmith and
Littlewood~\cite{Coppersmith87} using a
\emph{microscopic}  non-linear   model
of interacting degrees of freedom with  
 a huge number of dynamically
inequivalent  attractors. 
Related work  by 
 Tang, Wiesenfeld, Bak, Coppersmith
and Littlewood~\cite{Tang87} (henceforth TWBCL),
emphasized that the relatively
rare \emph{minimally stable} attractors 
of this model are nonetheless
those  preferably   selected by the dynamics.   
The sand pile model and SOC then
evolved~\cite{Bak96} from the 
analysis of the TWBCL model, with its
\emph{poised state}   being  conceptually 
similar   to a  minimally stable state.
 A  sand pile~\cite{Bak87}  reacts to small  
 disturbances  by releasing  avalanches
with a broad distribution of sizes, returning
 then to    its poised state described by the angle
of repose. 

While SOC deals with  the  \emph{stationary} fluctuations  of  
extended systems,   a wide class of systems 
is  manifestedly   non-stationary,  since the relevant
\emph{macroscopic} variables   slowly  change  in time at a 
\emph{decelerating} rate. This  implies   a   dependence of the data
on the initial time  and hence on the  \emph{age} of the system.
Relevant examples  are   
 spin glasses~\cite{Granberg88,Vincent91,Jonason98},
 the     evolution of   bacterial
cultures~\cite{Lenski94},    evolution in
rugged fitness landscapes~\cite{Sibani99a,Sibani97,Aranson97},
macro-evolution~\cite{Sibani95,Newman99b},
granular systems~\cite{Jaeger89,Josserand00}
and  Lennard-Jones glasses~\cite{Utz00}.
In spin glasses and glasses,  aging behavior 
 is  usually analyzed in terms
of   functions with two time arguments as e.g. magnetic correlations and 
linear response. Since at   `short' times $t<t_w$ these fulfill the
 Fluctuation Dissipation Theorem, one can infer   
that the  system  performs  equilibrium-like fluctuations
in this  regime~\cite{Andersson92}. For $t > t_w$ the FDT is broken and 
the non-stationary nature of  the dynamics becomes apparent. 
Intimately linked to   non-stationarity  
is  the reset capability of aging systems, 
i.e. the possibility of enhancing   the  rate of relaxation, 
thus `resetting' the system's apparent age to a smaller value
by   tweaking     parameters such as e.g. temperature and/or  magnetic
field~\cite{Granberg88,Vincent91,Jonason98}.

Below we use the TWBCL model, whose attractors
are explicitly known, for  a case study 
 of  the aging of   non-thermal
systems  with multiple metastability. 
Being particularly interested in 
 the connection  between the coarse grained dynamics 
and the attractor structure, we  
find it convenient to  consider  
the presence of two dynamical
regimes, pseudo-stationary for $t < t_w$ and non-stationary 
for $t > t_w$,  together with  the reset capability   as the   
central properties  of aging dynamics. 
These  properties are shared by   spin glasses and glasses, but not  
 by   e.g. the Bak-Sneppen~\cite{Bak93} evolution model,
whose macroscopic variable (the average fitness)
remains   constant in time. Nonetheless,  this   model
has other  interesting age dependent properties, as discussed
in Ref.~\cite{Boettcher97}. 
    
\section{The TWBCL model} 
In spite of its out most
simplicity, the TWBCL model with added noise
has interesting aging properties: 
The relevant macroscopic average,  
 here  called the degree of phase organization $\| x \|$, 
remains nearly constant on  scales $t<t_w$, 
and  the noise induced
fluctuations    are avalanches
(SOC like in two dimensions).  For $t>t_w$ 
a logarithmic decrease of $\| x \|$  
 becomes apparent,   revealing  that   the attractors visited 
become  more stable as the system ages. 
The  decay of  $\| x \|$    
can be reset  by a  change of  the elastic constant,
whereby the system is rejuvenated.  
All this behavior can   approximately be accounted for
 by  a mechanism previously
dubbed~\cite{Sibani93a} \emph{noise adaptation}, which  is
also present  in  the  dynamics of populations evolving  in
rugged NK landscapes~\cite{Sibani99a}.

Consider  $M$     `balls'   arranged in  an array
(linear or square) and coupled
to their  neighbors via springs
 with  elastic constant $K$.  
The balls are subject to 
friction,   to a force with a     sinusoidal 
spatial variation, and to a    
 series of    square   pulses
of  amplitude $E$.
In the  limit of high damping, large field  and weak elastic
coupling, the key dynamical  features 
are captured by  the  simple 
update rule~\cite{Coppersmith87}
 reproduced below (with  1D notation):  
\begin{eqnarray}
z_j(t) \quad & = & y_j(t) +  K \Delta ( y(t))_j + E + N_j(t), \nonumber \\
y_j(t+1) & = & \mbox{nint} (z_j(t)).
\label{autom}
\end{eqnarray}
Here,   $t$ is the   time in units 
of field  cycles,   $z_j$ is the coordinate of 
the $j$'th ball,   $\Delta$ is the lattice
Laplacian,     nint$(z)$   stands   for  the integer nearest  to 
 $z$ and $N_j$ is the noise applied at site $j$. 
\begin{figure}
\begin{center}
\includegraphics[width=8.5cm]{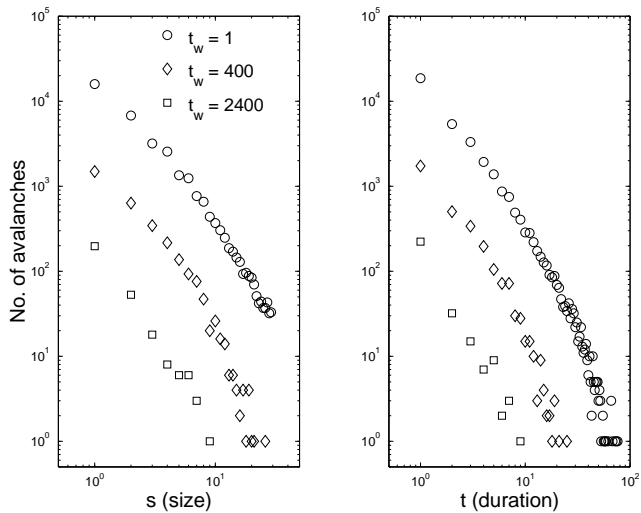}
\caption{\small Broad,  power-law like
distributions are observed for both   avalanche
sizes (left panel) and durations in a $30 \times 30$
model with $K =0.05$. The system  is subject to  
noise pulses drawn from an exponential distribution 
with average $a=0.01$ and the avalanches  are monitored 
through $100$  perturbation/relaxation cycles starting
at three different ages $t_w$.  The age is  here
the total number of cycles
the system has undergone before sampling the
statistics.   As $t_w$  increases
the avalanches  become  smaller and shorter.}
 \label{av_sizes}
\end{center}
\vspace{-0.5cm}
\end{figure} 
For $N=0$, integer valued   $E$
 and free or periodic boundary conditions,
the  attractor states
of Eq.~(\ref{autom}) 
satisfy~\cite{Tang87} $\mbox{nint} (K c ) = 0$, where 
 $c  = \Delta y  $ is   the curvature vector.    
The corresponding  coordinates then    fulfill  
\begin{equation}
\frac{-1}{2} \leq K c_j < \frac{1}{2} \quad j=1,  \ldots M.
\label{fixpoint}
\end{equation} 
The attractors\cite{foot}   
thus  lie   within an  \emph{attractor hypercube}  of
side length $1/K$ centered at 
the origin.   Their number  
is  ${\cal O}(1/K^M)$,  which is   huge
when,  for example,   $M\approx 200$ and $K=0.05$.

Noiseless relaxation of  an
initial state generally  leads to 
a \emph{phase organized
state}~\cite{Tang87}, i.e. a state  located at the corners of the
attractor hypercube. Such state  is \emph{minimally stable} against external
perturbations, as it  barely fulfills Eq.~(\ref{fixpoint}).
 The average    $\| x \| = M^{-1} \sum_i^M \mid x_i \mid$
is  always defined   and gauges, for  attractors,  
the  degree of (meta) stability $\| x \|$,
or, equivalently, the \emph{depth} $d = 1/2 - \| x \|$.
Minimally stable attractors have $d \approx 0$. 
 
\section*{Noisy relaxation properties}
\begin{figure}[h]
\begin{center}
\includegraphics[width=8.5cm]{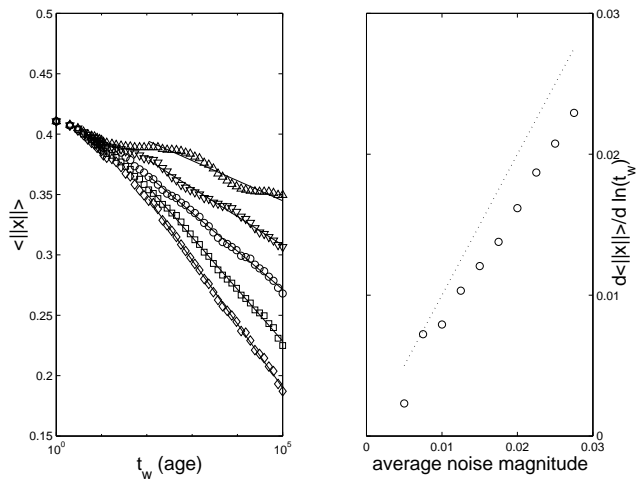}
\caption{\small Aging in a  $1000 \times 1$ model with    
$K =0.05$, randomly perturbed 
by white noise with exponentially
distributed magnitude of   $a$.  Each   curve in the left panel  
belongs  to a different value of $a$ and
is    the average over  ten independent trajectories, starting
from the same minimally stable state.  
After a short transient, the   decay is logarithmic 
(with a superimposed oscillation).
In the right panel, the  logarithmic  slopes are shown  versus
 $a$. A line of unit  slope is
included to guide the eye.} \label{TWBCL_noisy3}
\end{center}
\vspace{-0.5cm}
\end{figure}

We always start the noisy dynamics
 at a  phase organized state  selected under
noiseless conditions, and denote the  
 time elapsed under the influence of
noise by  $t_w$, the age of the system. 
As we anticipated, the   evolution has a first (pseudo)
stationary phase involving fluctuations    
among   meta-stable states  
of   the same depth.  
 On longer time scales  the average
 depth of the attractors visited increases logarithmically
through a series of jumps,  
 also denoted  \emph{macroscopic events}
or \emph{punctuations}. 
Crucially,  $t_w$ demarks the
boundary between short and long time
dynamics.   As shown by Fig.~\ref{TWBCL_noisy3},
our macroscopic average $\|x \|$  decreases  in
a  logarithmic fashion, 
apart from a superimposed oscillation which is   
most clearly visible for small noise amplitudes. Let $c$ be the
logarithmic slope of $\| x \|$, which is  shown in the
second panel of Fig.~\ref{TWBCL_noisy3}  and   
 assume that the observation
window extends from    $t_w$ to $t+t_w$.  
Since  $\ln(t + t_w) = \ln(t_w) + \ln(1 + t/t_w)$,
it follows that
\begin{equation}
\| x \|(t+t_w)  \approx \|x\|(t_w) - c \ln(1 + t/t_w).
\end{equation} 
Considering that $\ln(1 + t/t_w) \approx t/t_w$ 
and that   $c << 1$, we see that 
 $\|x\|$ does not change appreciably as long as   $t /t_w < 1$.
 Hence,  the  dynamics   appears stationary for $t < t_w$,
  as claimed. 
 
We can reach the same   conclusion by a second argument,
which will help us  to   connect  with the  
landscape structure of the problem:
By definition,  consecutive macroscopic
events always delimit the observation window 
during which the dynamics appears as  stationary. 
Secondly, as   we will show  later,
   the residence time $t_r$ 
 characterizing    the 
attractors   which are   first visited at time $t_w$   fulfills
(within an order of magnitude)
 \begin{equation}  
  t_r \approx  t_w. 
\label{fund}
\end{equation}
Hence  the dynamics 
appears  stationary  within the interval  $t < t_r \approx t_w$.
Interestingly, Eq.~(\ref{fund})  constitutes   the main   assumption of  
\emph{weakly broken ergodicity}~\cite{Bouchaud92},
a widely  used  scenario for 
complex system relaxation. The same equation    also describes
a  property of  diffusion   on 
\emph{hierarchical tree models}~\cite{Sibani89,Sibani97a,Hoffmann97},
models which reproduce many features of glassy relaxation.

Figure~\ref{av_sizes} illustrates the 
nature of the `short time'  avalanche  dynamics.  
The noise   used to produce the data
  consists of  a series of  `kicks'
of either sign, simultaneously applied
to each `ball' and  independently  drawn from 
 an  exponential distribution
with average $a$.
Relaxation  to a fixed point  is allowed between  
consecutive perturbations, and   the 
 avalanches consist of
 sets of  contiguous `balls'  simultaneously
in  motion. Their sizes are  
 defined as  the  largest
number of participating balls.
Both size  and duration
 are exponentially distributed in 1D,
and power-law distributed  in 2D,  as  
expected~\cite{Narayan94}.
The main   message  of  Fig.~\ref{av_sizes} is  
simply that avalanches are larger and last longer  in 
a young system (upper curve) than in an aged one
(lower curve). This is a first indication of 
decelerating dynamics.
 
Pulsed  noise can model    systems where the  
typical avalanche duration   and the length
of the noiseless periods are well separated~\cite{Datta00}. 
To bypass this restriction we now apply  the  noise 
`continuously', i.e. at each time step. 
The coarse grained  
time evolution in  state space can then be followed
by monitoring  $\| x \|(t_w)$.

Averaging suppresses
all   fast fluctuations together 
 with the spatial information,         
and $\|x \|(t_w)$ therefore consists   
 of constant  plateaus, punctuated
by rapid changes.   These mainly lead to deeper
attractors   and   stand 
  out as \emph{the} coarse grained 
dynamical  events  on long time scales. 
\begin{figure}
\begin{center}
\includegraphics[width=8.5cm]{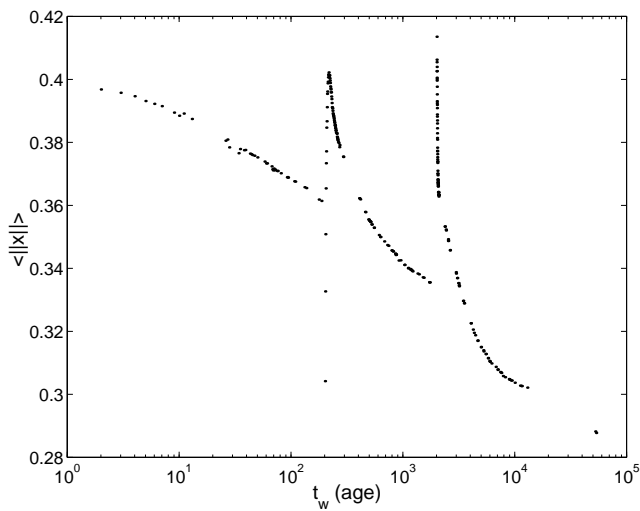}
\caption{Changing the elastic
constant from $K=0.03$ to $K=0.05$ and again to 
$K=0.07$  produces  the  
`resets'   seen   at times
$2\cdot 10^3$ and $2\cdot 10^4$. The data are  
averages  over $20$ different trajectories of  a  linear array
of  $1000$ `balls'. The noise magnitude is exponentially
distributed with  average $a=0.015$.} \label{TWBCL_noisy4}
\end{center}
\vspace{-0.5cm}
\end{figure}
Complementary  information  is  obtained by  
 averaging    $\| x \|$ over independent   noise histories.
The resulting   smooth function, $\langle \|x \| \rangle (t_w)$
was   studied  for  both 1D  
and  2D systems, using pulsed as well as  continuous
noise. Since  $\langle \| x \| \rangle$  is  rather 
 insensitive to  the   dimension,  
we mainly studied it for  1D models, which are faster  to 
simulate.  
The left panel of Fig.~\ref{TWBCL_noisy3} shows the time evolution
of  $\langle \|x \| \rangle (t_w)$ for different values of the noise magnitude
$a$.  The negative logarithmic slope of the plots  has
a  linear relationship to the amplitude $a$ which is shown in
 the right panel of the same figure.

The  age-reset   
is  induced in our  model by changing the elastic constant, 
which is analogous to  changing the   magnetic field~\cite{Djurberg95} 
in spin glass systems.   
Increasing  $K$   reduces  the size
of the  attractor    hypercube and  concomitantly 
reduces the depth  of the current state.
As a consequence    $ \langle \| x\| \rangle (t_w)$ is  reset
to an earlier  (and larger) value,  as shown in     
Fig.~\ref{TWBCL_noisy4}. Decreasing $K$ has the
 effect of swelling  the attractor hypercube
whence  $\langle \| x \|\rangle $ quickly drops.

To further clarify the connection between
the reset effect and  the attractor geometry,
 we  consider
  \emph{bounded noise}~\cite{Sibani93a} drawn 
\begin{figure}
\begin{center}
\includegraphics[width=8.5cm]{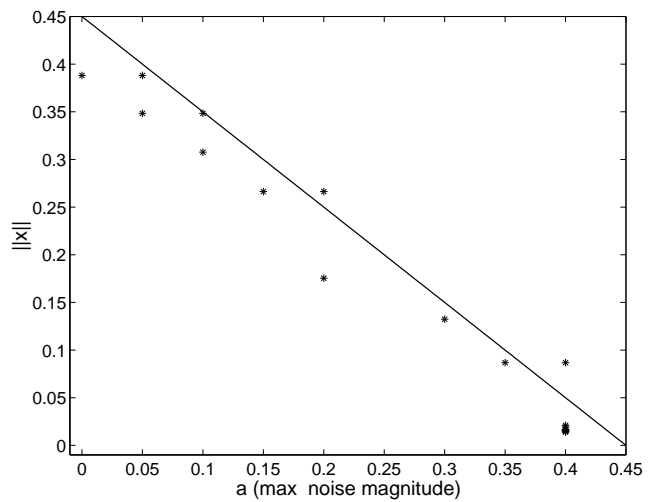}
\caption{\small A $1000 \times 1 $  model with     
 $K =0.05$ is    perturbed through a few thousand
updates by   noise of bounded variation
   $[-a,a]$ and then allowed to fully  relax, reaching
the  $\| x \|$ values which are 
 plotted  (as stars) versus $a$ for five    independent  noise sequences. 
In addition, the line $y = 0.45 - a$ is drawn as a guide
to the eye. Since   $\| x\|$ decreases almost proportionally to $a$,  
the \emph{least stable} among the attractors   
surviving the noise  are  those dynamically selected 
with high probability.}
 \label{TWBCL_noisy2}
\end{center}
\vspace{-0.5cm}
\end{figure}
from a uniform distribution  supported  
in the finite interval   $[-a, a]$.
  Equation~(\ref{autom}) then implies that only  
states  fulfilling  $\max_j  \{1/2 - \mid x_j \mid \} > a $ 
survive as  \emph{exact} fixed points of the equations of motion.
 The  corner states of an  attractor hypercube  of side length $(1-2a)/K$   
are still  minimally stable,   in the  generalized sense that 
any infinitesimal  \emph{increase}  of the noise amplitude   destroys
their stability. 
Figure~\ref{TWBCL_noisy2} demonstrates   that 
a perturbation  of magnitude 
$a$  permanently leaves   the
system  in an attractor of  depth   
$ \approx 0.45 - a$. Hence  the 
minimally stable states are dynamically selected. 
  
For   exponentially distributed noise
of average magnitude  $a$ and on a time scale
$t_w$,   the kicks    
normally  fall  within  the  range 
$r_t \approx   a \ln t_w $.  
Hence we expect that the trajectories will 
typically be located at the corners of a hypercube
of side length  
$ 2 \langle \| x \| \rangle \approx (1  - 2 a \ln t_w)$.  This
is in reasonable agreement 
with the behavior depicted    in the right panel of  
Fig.~\ref{TWBCL_noisy3} and explains why 
 even a  modest  shrinking
of  hypercube produces a sizable  reset. 
Since the 
attractors typically discovered  
on a given  time scale are  the shallowest among those
available, we expect that a    \emph{record} in the
sequence of noise kicks will  likely suffice to produce a 
macroscopic event, a  feature   previously dubbed~\emph{noise}
\emph{adaptation}~\cite{Sibani93a}. 

\begin{figure}[t]
\begin{center}
\includegraphics[width=8.5cm]{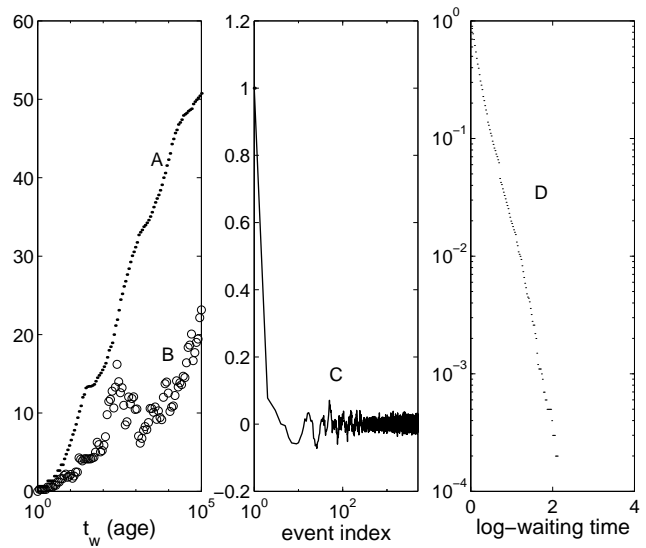}
\caption{An array of 1000 `balls' with  elastic
constant $K=0.05$  is  
perturbed  by noise  `kicks' of random  sign  
and   magnitude exponentially distributed with average  
$a=0.015$. The noise is uncorrelated in time and space.
We considered 200  independent  trajectories,  all starting 
from the same minimally stable state. 
An `event' is defined as  
the achievement of a state of lower $\|x\|$.  
A and B:  average   and variance of the
number of events observed within  time $t_w$.
C: autocorrelation function   $C_\Delta(k)$ 
of the log-waiting  times $\Delta_k = \ln(t_k/t_{k-1})$. 
D: distribution of the $\Delta_k$.}
 \label{portrait}
\end{center}
\vspace{-0.5cm}
\end{figure}

If macroscopic events are induced by  noise
records,     their number  $n_e(t_w)$
during time  $t$  is 
a \emph{log-Poisson} process~\cite{Sibani93a}. 
As a consequence,  if  $t_k$ denotes the   time of the $k$'th event,
the quantities 
$\Delta_k = \ln (t_k/t_{k-1})$  are   
statistically independent and have the    common   distribution 
Prob$(\Delta > x) = \exp(- \lambda x)$, 
for some positive $\lambda$.  Secondly,  
the  average number of events  grows  as 
$  \langle  n_e \rangle (t_w)= \lambda \ln t_w$. 
From Fig.~\ref{portrait} we see  that  the       
actual statistics of macroscopic events  
resembles a   log-Poisson statistics in 
the shape of $ \langle  n_e \rangle $ (plot A) and in  the
fact that the log-waiting times have 
very short correlations (plot C)   and  an  exponential 
distribution (plot D).   
 
Crucially, Fig.~\ref{portrait} D  
shows  that the event times $t_k$  
approximately make up  a geometrical series. 
Hence,  in a system of  age $t_w$, most   
time is  spent   in  the neighborhood of the  
 last attractor visited. Therefore, 
 the residence time in a neighborhood
of the $k$'th attractor discovered is   
$t_r = t_{k} - t_{k-1} \approx t_{k} = t_w$, 
as  anticipated in Eq.~(\ref{fund}).  
 
\indent 
\section{Summary and conclusion}
On short time  scales  the space resolved  dynamics of the 
TWBCL model  can be described in terms of avalanches 
 having, in two spatial dimensions,   SOC-like character.
On longer time scales  the 
applied noise  pushes   the system into
gradually more stable attractors.  As a consequence, the degree  of phase
organization $\|x\|$ decreases logarithmically in time. 
This  differs from the sand pile model and  is reminiscent of the
logarithmic relaxation of the angle of repose~\cite{Jaeger89}   
observed in  actual  sand piles subject to vibration. 
We have argued that  the  aging of the TWBCL model
is similar to that of   e.g. spin glasses in two 
important respects: i) the boundary between quasi-stationary and non-stationary 
dynamics is given by $t_w$, the time elapsed from the initial 
quench, and ii) the dynamics is resettable.
The coarse grained aging dynamics is characterized 
by  a series of `macroscopic events' leading to 
gradually deeper attractors. The statistics of these events is
approximately log-Poisson, an indication that the
events themselves are strongly correlated with records
in the  history of noise. During   
noise adaptation~\cite{Sibani93a}   
the attractors first visited on a  time scale $t_w$  typically 
trap the  trajectories for  time   $t_w$, as 
 assumed in   weakly broken 
ergodicity~\cite{Bouchaud92}. The same statistics
is also present in the   dynamics of a population
of `agents'  evolving in   NK  fitness landscapes 
with multiple optima~\cite{Sibani99a}. If  one views
evolution as a search in a 
fitness landscape with multiple optima,   
stress-induced hyper mutation~\cite{Bridges97} 
following a change of nutrient type or concentration 
appears as the  biological counterpart 
of a reset.  
Thus,  a range of complex non-stationary 
phenomena   can be qualitatively understood  by invoking  
marginal stability and noise adaptation as
selection mechanisms for metastable attractors. 

\noindent {\bf Acknowledgments}:
P.S. has been supported by the Danish Research
Council.

\bibliographystyle{apssty}

\bibliography{thesis}

\begin{thebibliography}{10}
\bibitem{foot}
In terms of the original variable $y$, these attractors are sliding
states for non-zero $E$ and fixed points otherwise. Since actual value of $E$ 
is immaterial for our purposes,   we use $E=0$.
\end{thebibliography}

\end{document}